\documentclass[sigconf]{acmart}
\usepackage{epigraph} 
\usepackage{tabularx}
\usepackage{booktabs}
\usepackage{caption}
\usepackage{array}     
\usepackage{ragged2e} 
\usepackage{xurl}
\newcolumntype{L}{>{\RaggedRight\arraybackslash}X} 
\newcolumntype{M}{>{\RaggedRight\arraybackslash}p{6cm}}
\AtBeginDocument{%
  }

\setcopyright{none}
\acmDOI{}
\acmISBN{}

\acmConference[LIMITS '26]{12th Workshop on Computing within Limits}{Month xx--xx, 2026}{Online}
\acmISBN{978-1-4503-XXXX-X/2018/06}

\begin{document}

\title{The LIMITS of Time}

\author{B Biira}
\affiliation{%
  \institution{University of Washington}
  \city{Seattle}
  \country{USA}}
\email{bbiira@uw.edu}

\author{Amelia Lee Doğan}
\affiliation{%
  \institution{University of Washington}
  \city{Seattle}
  \country{USA}}
\email{dogan@uw.edu}

\renewcommand{\shortauthors}{Biira and Doğan}




\begin{abstract}
The LIMITS community was founded to foster conversations that move away from growth‑oriented visions and values in computing toward a focus on long‑term well‑being. This orientation, we argue, inherently engages questions of time and temporality. Prior work has shown that temporal frameworks shape how futures are imagined, which problems are understood to be worth attending to, and which solutions or alternatives are pursued. We begin this paper with author observations of time in their lived experience, and then extend these observations to the LIMITS community. Through a systematic literature review of the last decade of LIMITS scholarship, we identify ways that explicit attention to how concepts of time and temporality are understood would enrich Limits scholarship. Within the LIMITS scholarship that does engage with time, we identify five recurring types of temporal engagement: computing time, methodological and design time, politics and ethics of time, biological and ecological time, and afterlife and waste time. Together, these engagement types highlight how implicit assumptions about time are embedded across research practices, design approaches, and accounts of technological impact within LIMITS work. We discuss these findings in relation to cross‑disciplinary scholarship that takes time as an analytic concern and consider how these patterns point to a broader need for more explicit, plural, and situated engagements with time in the LIMITS community, and why this matters for the community's commitments.

\end{abstract}

\maketitle 

\section{Introduction}
\epigraph{the muslim man who abandons \\ his car at the traffic light drops\\ to his knees at the call of the azan} {Fatimah Asghar, \textit{If They Should Come for Us}}

\epigraph{\quad To the swinging and the ringing\\
\quad \hspace{0.1cm} Of the bells, bells, bells, \\
\quad Of the bells, bells, bells, bells, \\
\qquad Bells, bells, bells— \\
To the rhyming and the chiming of the bells!}
{Edgar Allan Poe, \textit{The Bells}}

These are poetry excerpts to some of the most universal timekeeping sounds our readers might have heard. The azan or adhan often recited from a minaret of a mosque may call people to prayer as described by Asghar. For Poe, the bells may have been church bells rhyming and chiming to celebrate a joyous occasion like a marriage or simply chiming on the hour. Often these assumptions about time, fade into the background sounds, even though a temporal assumption is one of the ``fundamental analytical facets of relevance to both the use and design of computer technologies'' \cite{rahm-skageby_hci_2022}. 

These sounds, and other timekeeping devices we use, help shape our very perception of how we move through the world. As many have written about \cite{rifkin_beyond_2017, kafer_after_2021, mbiti_african_1970}, the Western ``clock time'' of twenty-four hours per day in a 365-day year is not the only way to keep time or experience it. Conversations about time span multiple scholarly communities, where temporality is taken seriously given its entanglement with environmental processes \cite{Edensor2020, may_timespace_2003,marquardt_making_2021, gokmenoglu_temporality_2022,fent_political_2020}. Within the disciplinarily diverse LIMITS community, scholarship often speaks to tensions and constraints that are implicitly tied to questions of time, even when temporality is not addressed directly or systematically. Time is not only embedded in the issues under study by the LIMITS community, but also in the conditions that have produced them. Ideas of speed, progress, and efficiency in computing, for instance, have shaped how we arrive at the present, often under forms of justification that remain largely unquestioned \cite{pargman_limits_nodate}. With that, our interest to foreground temporal assumptions stems from how they have operated as taken-for-granted rationales for processes that, while making possible growth and other generally accepted to be beneficial contributions to inhabitants of the planet, also define some of the very limits now being confronted.

At the same time, engagements with temporality within LIMITS work, while often implicit, sometimes emerge through reflexive and recounting practices across both professional and personal contexts \cite{penzenstadler_leverage_2020,pargman2023exploring}. Our motivations started from reflections and recounts from both authors, informed in part by experiences in a graduate course that provided the opportunity to engage at length with LIMITS scholarship, as well as experiences outside the classroom that have shaped how we understand and engage with the community’s commitments. These motivations and reflections are positioned here before turning to the research questions that guide the rest of the paper. Building on a tradition of personal reflection and recounting in prior LIMITS work \cite{penzenstadler_pursuit_2021,penzenstadler_leverage_2020,mayhew_materiality_2021}, we turn to one author's observations and experiences in their attempt to reconnect with other ways of time.

\section{Author Reflection}
Several years ago, I\footnote{Amelia, the second author} bought a traditional lunar calendar made especially by other diasporic Chinese artists\footnote{\url{https://sites.google.com/view/lunarcalendar/home}}. The project was aptly named ``Chinese Diaspora Lunar Calendar'' with an explicit purpose of trying to help diaspora members reconnect with traditional lunisolar time keeping. As the artists, Juniper Wong, Chloe Wang, and Camey Yeh, specifically write in their introduction to the calendar, ``We have created two iterations of a printed calendar, for the years of the Water Tiger and Water Rabbit, that center the lunar months, solar terms, and festivals that comprise the traditional Chinese yearly cycle. To make these accessible in the context of dominant, colonized ways of time-keeping, we include corresponding Gregorian dates and accompanying interpretation. Looking to traditional knowledges and rhythms is part of an effort to revive lost and latent elements of our roots among the branches of diaspora.'' The project was not a complete rejection of the Gregorian calendar that I live on, but was trying to introduce a different seasonality and rhythm that my ancestors have lived with for quite literally thousands of years.


This coexistence can be seen in the physical layout of the calendar of the ninth lunar month from the calendar for the year 4270. The ninth month spread from the calendar for the year 4270 includes a wood block print by Jennifer Zee\footnote{\url{https://www.ginkgozee.com/}} of tigers and ginkgo leaves, echoing the year's designation as the Year of the Water Tiger. Chinese years are designated using one of the five elements (wood, fire, earth, metal, and water) and one of the twelve zodiac animals to create a sixty-year cycle. The lower part of the calendar includes a traditional calendar layout from the ninth month of the water tiger year 4270 (Gregorian year 2022), but there are Gregorian dates aligned to each of the calendar days, such as the month beginning on September 26, and the annotation for October 1 corresponds with the sixth day of the calendar. There are also both Western and Chinese holidays affixed to the calendar, including the Double Ninth Festival, which is celebrated on the ninth day of the ninth month. The festival often includes celebrations to clean ancestral gravesites. In a digital booklet\footnote{\url{https://sites.google.com/view/lunarcalendar/digital-booklet}} meant to help familiarize diaspora with the calendar, the artists include an eighth-century poem (Double Ninth, Remembering my Shandong Brothers) by (Wang Wei). The accompanying booklet helps to orient calendar users to not only the time differences but also the cultural and ancestral knowledge that were developed in conjunction with the differing temporal frame. In this calendar, as a melding of different time systems, it also includes Indigenous Peoples' Day, an American holiday on the second Monday of October, started as a counter-celebration to Columbus Day, to celebrate Indigenous American history instead of the Italian explorer's bloody legacy.\textit{ }

The calendar also has the solar terms. Twenty-four solar terms make up subdivisions of the Chinese lunar calendar based on $15^{\circ}$ intervals of the Earth's orbit on the Sun. It was developed during the ancient Chinese agrarian society \cite{zhou_following_2023}. For example, the calendar includes notations of the start of Cold Dew and Frost Descent. Cold Dew represents how the weather has turned chillier ``the water vapor in the air is condensed to form dew at night'' \cite{zhou_following_2023}. The digital booklet with the calendar offers practical advice for the user to ``trade cotton for wool, and iced drinks for warm porridge and tea, to nourish the body's heat from inside and out. Turn off the air conditioning; protect your internal organs.'' These seasonal changes help demonstrate how the calendar speaks to Chinese cosmologies of medicine and food, not just temporal practices. Frost Descent is the solar term of autumn as ``dew condenses into frost, and the colder weather can be felt everywhere\ldots During this transition time from autumn to winter, Chinese local food customs are distinctive\ldots Persimmons and apples are often among the dietary supplements \cite{zhou_following_2023}.

\begin{figure}
    \centering
    \includegraphics[width=\linewidth]{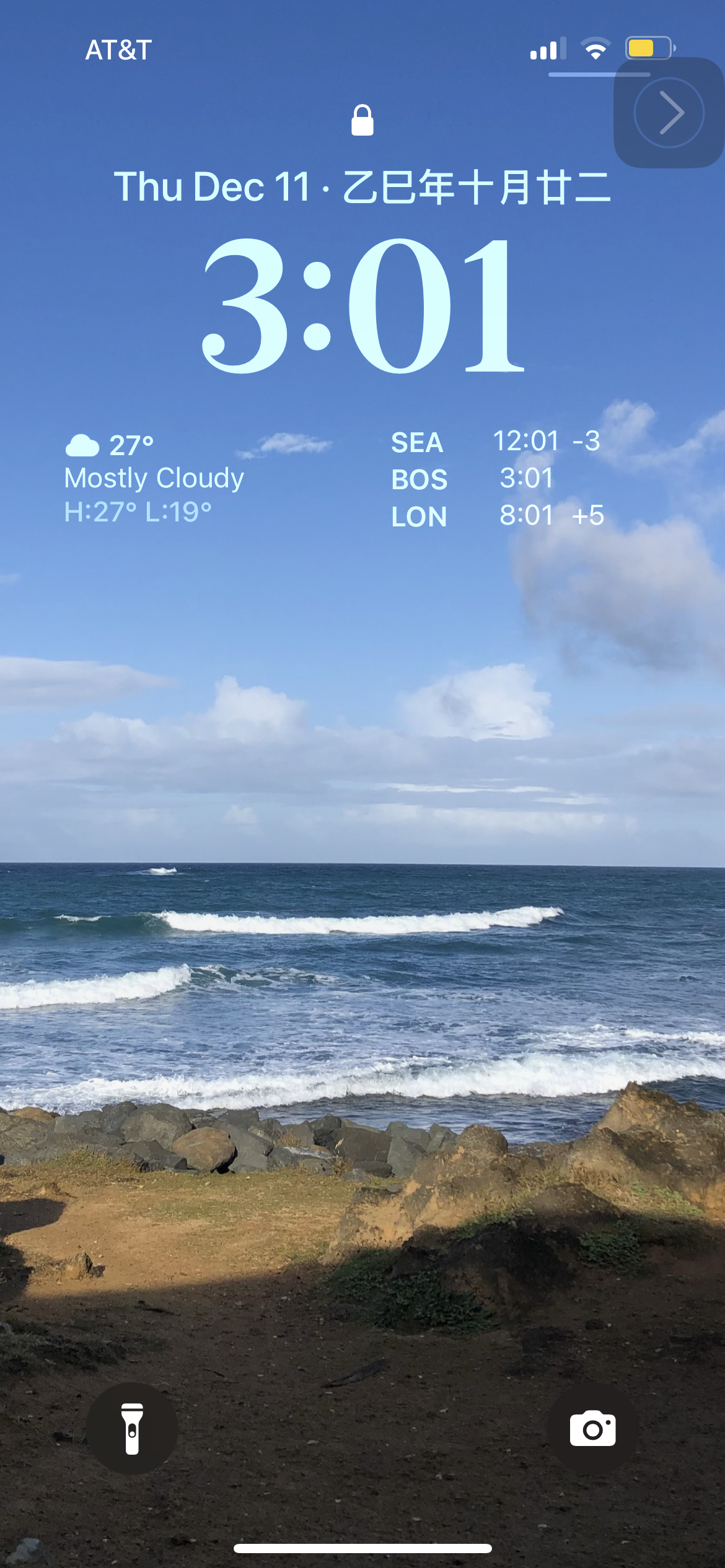}
    \caption{Phone lock screen showing Gregorian calendar date and traditional lunar date.}
    \Description{Screenshot of iPhone lock screen with a background of a wave crashing on the beach. The top of the lock screen includes the text Thu Dec 11 with Chinese characters showing the date in the traditional lunar calendar.}
    \label{fig:lunar-phone}
\end{figure}

As I used the calendar throughout the year, I found my time a bit discombobulated with trying to understand this calendar (re-)introduced to my life. There were holidays I had celebrated growing up, but I would often sense them coming and Google what day they were coming up. At the same time, the calendar did not necessarily integrate into my everyday life, which is spent staring at my iCal that uses the Gregorian calendar. In the digital world, on Apple and Google calendar products, it is possible to add the Chinese calendar to the default Gregorian calendar. I did this on my phone home screen as shown in Figure \ref{fig:lunar-phone}. On the top of the home screen, I can see the date is December 11 in the Gregorian calendar and also the 22nd day of the tenth month of the Wood Snake year. In this way, I tried to bring this other temporal framework into the everyday through both a wall calendar in my room and on my phone. It's not a singular way of being or of telling time but is an ongoing experiment using digital and physical tools to lean into ancestral timekeeping, cosmology, and practices. 

\section{Research Questions}
As authors who have considered and attempted to live with other temporal frames of reference, we were curious if our fellow researchers in the LIMITS community have attempted to articulate or consider temporal frames within their own work. While LIMITS work is often concerned with long-term thinking, sustainability, and alternative futures, time and temporality are not consistently centered. In many of the questions the community engages, temporal considerations remain implicit and are often taken for granted, leaving unclear how they are conceptualized within LIMITS scholarship. To explore this, we asked:

 \begin{itemize}
     \item \textbf{What temporal assumptions are present in LIMITS scholarship? }
     \item \textbf{How might conceptions of time in LIMITS scholarship shape the community's thinking and assumptions about what is possible?}
 \end{itemize}

We first explored various approaches to time beyond Western approaches to ``clock time'' \cite{rapp_introduction_2022}. We then conducted a systematic literature review of a decade of work published in the LIMITS workshop to examine how time and temporal frameworks are taken up across the community’s contributions. Through this analysis, we identified and present five recurring ways in which time is engaged, often in implicit ways. Finally, we reflect across this literature and the personal experience introduced earlier to consider how LIMITS might more explicitly engage temporal frameworks within their work. In doing so, we highlight how attending to time directly can deepen the community's approach to sustainability, futures, and the design of computing systems.

\section{Some Notes on Time} \label{related-work}
There are works on how time outside of the Western academic tradition is conceived as non-linear \cite{rifkin_beyond_2017}. These approaches contrast with the chrononormative approach to ``clock time'' that orients toward maximizing labor and linear life events \cite{hartmann_normative_2019}. This includes theorizations such as ``crip time,'' coming from disability studies. Crip time can be conceived as literally, the very different time that disabled people lived with managing their disability or shortened life expectancy, but also a theoretical approach to temporality \cite{samuels_six_2017}. As Allison Kafer \cite{kafer_after_2021} writes, ``Maybe we should think less of what crip time is and more of what crip time does, thinking beyond specific speeds, toward as yet unimagined imaginaries. What are the temporalities that unfold beyond, away from, askance of productivity, capacity, self-sufficiency, independence, achievement?''  Thinking of time in different ways can open up new possibilities and speculation within research. We seek to explore how Western time, ``settler time,'' or ``clock time'' has become the dominant way of configuring our lives and research. 

Assumptions about time have already been explored in various domains related to human-computer interaction (HCI) \cite{wiberg_time_2021}. Rahm-Skågeby and Rahm \cite{rahm-skageby_hci_2022} have talked about HCI's focus on short-term interactions, and human-only perspectives are insufficient to address Anthropocene-scale problems. By not declaring our assumptions about time within sustainable HCI research, and the broader HCI field, we risk conflicts in the field about what we mean by sustainability \cite{laurell_thorslund_exploring_2025}. Relatedly, other work has sought to reorient professional development in computing and related fields through reflexive practices around processes of learning and teaching, and how these shape sustainability and imagined futures \cite{pargman_experimenting_2019, mann_regenerative_2018, penzenstadler_leverage_2020}.  Wieczorek et al. \cite{wieczorek2026whose} use \textit{temporal arrangements} to describe the ways time is shaped and allocated across infrastructures in ways that unevenly distribute burdens and demands. Other research has tried to push the field to consider non-human times such as daily cycles, monthly cycles, seasonal life cycles of plants and animals \cite{griffiths_time_2023}. And other work, such as research-through-design has expanded on methods to surface and encourage reflexive practice with respect to concepts of non-human, non-anthropocentric, non-linear temporalities \cite{zohar_when_2024}.

Outside of HCI, many humanistic research traditions have also critiqued time. For example, environmental humanities critiques anthropocentrism by using the concept of deep time, drawing from climate science, to place human history within the larger planetary setting \cite{irvine_deep_2014}.  Yet, even climate science finds it troubling to communicate the concepts of change over the millennia of deep time \cite{caseldine_conceptions_2012}. Rather, environmental humanities urges us to understand and hold the tensions of multiple temporalities \cite{senda-cook_engaging_2023}. Below, we seek to explore various other temporalities. 

Drawing on a decolonial tradition, various scholars have written about how Indigenous and non-Western cultures conceptualize time and temporality beyond linear time. For instance, Iparraguirre draws a distinction between time (as the becoming in itself) and temporality (human interpretation shaped by cultural context) \cite{iparraguirre_time_2015}. Western, linear conceptions of time represent a form of hegemonic temporality, while cultures like the Mocoví Indigenous group in Argentina, operate within non-linear, cyclical, or immediate temporalities that blend ``historic and mythical narratives along with everyday personal experiences'' outside bureaucratic time structures that characterize hegemonic temporalities. In another case, Mbiti \cite{mbiti_african_1970} writes about how traditional African thought rejects the Western linear time and instead views time as two-dimensional, moving backward and forward. He characterizes African traditional temporal frameworks as the immediate present (Sasa) and the vast, mystical past (Zamani), experienced as the direction of time moving from Sasa to Zamani.  Sâkihitowin Awâsis \cite{awasis_anishinaabe_2020} writes about how Anishinaabe ways of living embrace multiple, non-linear and intergenerational temporalities that are relational and ``enmeshed with the land.'' For example, the concept of \textit{aanikoobijigan} refers to both ancestor and descendant, demonstrating the intergenerational spiral of time where both past and future collide into the present. 

We give these examples to animate various other traditions the LIMITS community could engage with when engaging with time. 

\section{Research Approach}

We performed initial screening to identify temporal engagement, followed by focused coding and reflexive annotation of the included papers to conduct a systematic literature review of the LIMITS literature. We detail our collection, screening, and analysis procedures here.

\subsection{Data Sources}

Our working corpus includes all papers (n = 159) from the LIMITS workshop proceedings published between 2015 and 2025. Papers were sourced from the LIMITS webpage (2015, 2024, 2025) \url{https://computingwithinlimits.org/}, the ACM digital library LIMITS archive (2016-2019) \url{https://dl.acm.org/conference/limits/proceedings},the ACM digital library IC4S archive (2020) \url{https://dl.acm.org/doi/proceedings/10.1145/3401335}, and the LIMITS Pubpub (2021-2023) \url{https://limits.pubpub.org/}. 

\subsection{Inclusion and Exclusion Criteria}

Our initial criteria required time or temporality to be explicitly discussed as an essential and integral component of a paper's contribution. And for screening, authors split corpus among themselves and performed an initial keyword search across the entire corpus using the following queries:

\begin{verbatim} "time", "temp", "past", "future", "speed", "slow", OR "fast" \end{verbatim}

\begin{figure}
    \centering
    \includegraphics[width=\linewidth]{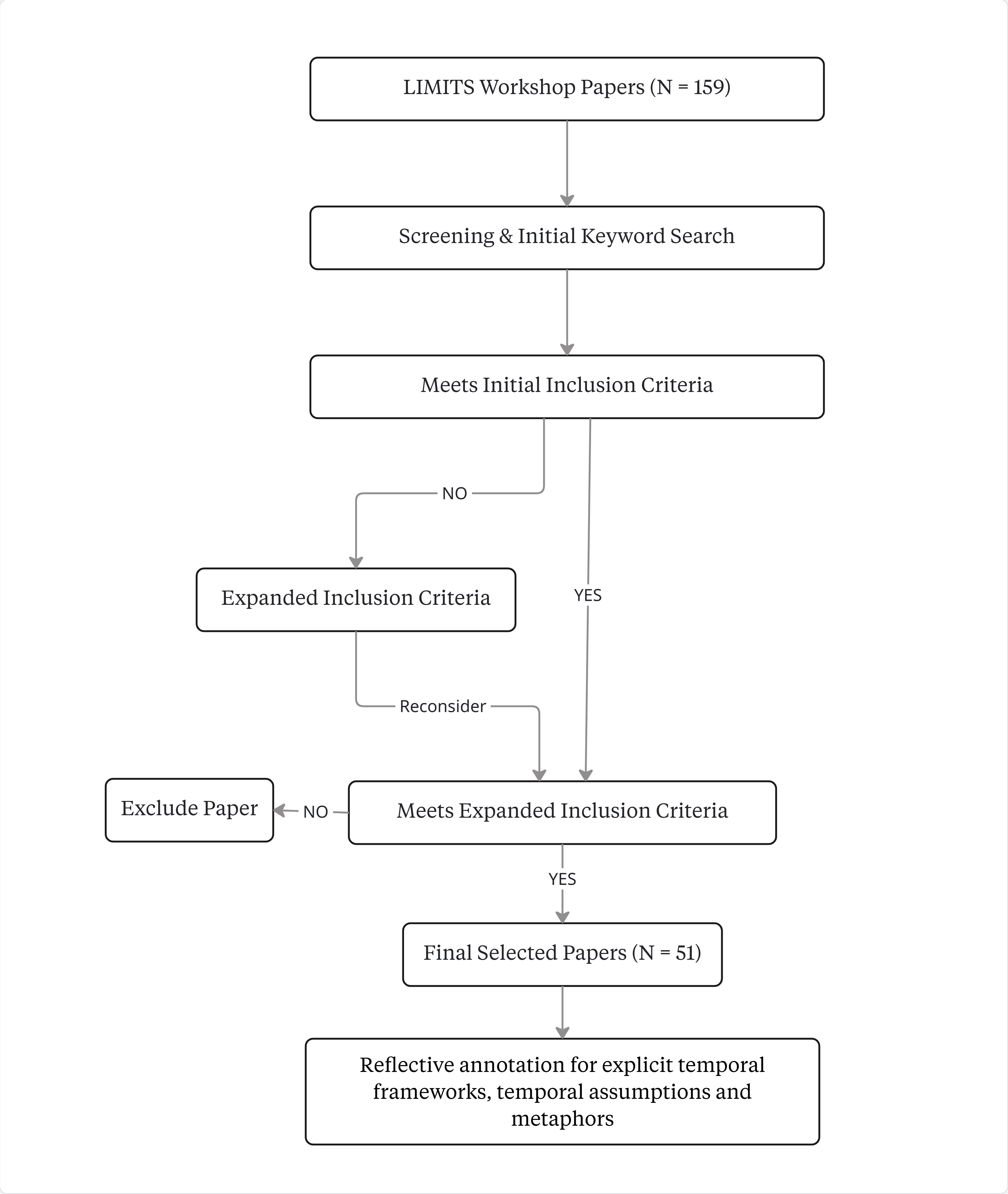}
    \caption{Flowchart of the review process}
    \Description{Review process started with the sourcing LIMITS workshop papers from years 2015 to 2015. Papers first underwent screening and initial keyword search to see if they met initial inclusion criteria. Initial inclusion criteria was adjusted and papers screen against expanded inclusion criteria. Papers that satisfied this expanded inclusion criteria became the final selected papers; those that did not were excluded. The final selected papers then proceeded to reflective annotation for explicit temporal frameworks, temporal assumptions and metaphors.}
    \label{fig:methodflow}
\end{figure}

Early in the screening process, we realized our initial inclusion criteria yielded few papers. We reconvened and adjusted our approach, expanding inclusion criteria to capture papers dealing with implicit or assumed temporal issues. This adjustment resulted in 51 papers that engaged with time, temporal assumptions, or time-related metaphors somehow, either implicitly or explicitly (see Figure \ref{fig:methodflow}. )

Contributions that did not meet the expanded inclusion criteria, mostly papers whose contribution did not hinge on or substantially discuss any temporal concept, assumption, or metaphor, were excluded.

\subsection{Reflexive Annotation}

For the 51 included papers, the first author annotated each paper for explicit temporal frameworks, temporal assumptions, absences, and metaphors. Reflexive notes were taken asking \textit{what does this temporal assumption do with respect to the paper's contribution or argument?} We present our findings in the section below.  

\section{Time in LIMITS}

We present here our findings on conceptualizations of time within the LIMITS Workshop literature spanning the last decade (see Table \ref{tab:table1}). 

\begin{table*}[t!]
\centering
\begin{tabularx}{\textwidth}{@{} M L @{}}
\toprule
\textbf{Type of Engagement} & \textbf{Description} \\
\midrule
{Computing Temporality} & Cultural practices and assumptions of time within computing practice, education, and research. \\
\midrule
{Methodological and Design Temporality} & Methods, design approaches, or ways of engaging with computing that propose alternative forms of time for the computing community, users, or systems. \\
\midrule
{Politics and Ethics of Time} & Political or ethical dimensions of time in computing, and engaged with aspects of power, control, labor, or inequality related to time. \\
\midrule
{Biological and Ecological Time} & Temporal aspects inherent to non-human living creatures or natural systems, such as animal growth cycles, ecological clocks, and biological rhythms on Earth. \\
\midrule
{Afterlife and Waste Temporality} & Focuses on the afterlife and end-of-life in computing, including waste, repair, maintenance, decay, obsolescence, or the digital afterlife. \\
\bottomrule
\end{tabularx}
\caption{Type of temporal engagement within LIMITS 2015-2025}
 \Description{Table presents five engagement types within LIMITS literature over the past decade 2015 to 2025e. The first column names the type of engagement, and the second column provides the brief description of the engagement type.}
\label{tab:table1}
\end{table*}

\subsection{Computing Temporality}

Work under computing temporality captured, within their contributions,  how the culture of computing as field constructs time. This spanned from professional development pedagogy and education \cite{soden_photovoltaic_2022, mayhew_materiality_2021}, to wider professional community norms \cite{penzenstadler_leverage_2020, pargman_limits_nodate}.  Mayhew \cite{mayhew_materiality_2021} identifies, through metaphor, how Computer Science education creates an ahistorical narrative, noting that \textit{the idea that code is magic and the creation of something out nothing... sets up this narrative that computing is immaterial.} In treating code as instantaneous magic, the discipline effectively erases the long, extractive timeline of hardware production. This, they write, creates a pedagogical gap where students \textit{don't get to see how it’s all a social construction}, preventing them from critiquing the trajectory that led to the current climate crisis \cite{mayhew_materiality_2021}. Soden et al. \cite{soden_photovoltaic_2022} turned to structural issues of time in computing culture. They write about how research timelines are incompatible with sustainable community engagement, which requires trust-building. This temporal pressure reflects the fast-moving culture within the computing industry and education, leaving little room for ethical reflection. Hendry identifies a similar issue in research design in his review of LIMITS papers on food. He notes that, \textit{seven LIMITS projects by and large do not make explicit commitments to time}, revealing a disconnect between the community's long-term rhetorical goals and its short-term methodological commitments\cite{hendry_transition_2021}.

\subsection{Temporality in Methods and Design}

Contributions in this category proposed methods and design approaches that depended on engagement with time. These contributions were either orientated towards research and design contexts approaches, though proposed applications extended across diverse fields. Bhardwaj et al., for instance, related design methodological approaches beyond computing to climate policy by pointing to ways temporal distance shapes perceptions of urgency and responsibility \cite{bhardwaj_limits_2024}  and how that extends to design of tools and other artifacts used for both meaning-making and decision-making in climate-related contexts \cite{bhardwaj_pathways_2023}. Others draw on historical and alternative knowledge traditions to inform present practice. Colliaux et al., \cite{colliaux_computational_2022} turned to the agricultural framework of French Intensive Method to demonstrate ways techniques can deliberately work against seasonal constraints to enable year-round production, thereby proposing a design approach that reconfigures temporal assumptions in practice. Still other contributions contribute methodological innovation directed to the LIMITS community itself. Eriksson and Pargman \cite{eriksson_meeting_2018}, for example, proposed counterfactual history as a design method to shift focus from forecasting or extrapolating the present forward. to rewriting or the altering the past to produce alternative presents. Rejection of forecasting in favor of backcasting was also noted. Pirson et al. \cite{pirson_notitle_2023} described backcasting as \textit{``working backwards from the end-point vision to the present''} specifically for changes occurring \textit{``over 20 to 100 years''}. As a method, this approach changes the function of time in design with the present forced to submit to the constraints of a future with finite resources. Bhardwaj et al. \cite{bhardwaj_limits_2024} highlighted the cognitive difficulty of grasping long-term climate impacts. They argue that \textit{``designing decision tools to aid in the traversal of psychological distance can be a way to compress time, space, and complexity.''} Their speculative design approach allows \textit{``time to be 'bent'... because they are allowed to presently feel the future.''} The metaphor used to describe how the method works suggests time is thought of as a gap to be bridged and as linear.

\subsection{Politics and Ethics of Time}

A set of contributions engaged with time in relation to control, organization, and coordination. Hilty \cite{hilty_computing_nodate} introduced the concept of \textit{temporal flexibilization of demand},  framing computing demand as something that can be coordinated with the fluctuating availability of renewable energy (e.g., solar peaks). With that, they positioned time as a mechanism for organizing computational activity and moderating resource use, challenging on-demand perspectives while raising ethical questions about who bears the burden of flexibility. Other contributions challenge neutrality of temporal aspects such as speed and waiting. Jacquet and Luxey-Bitri \cite{jacquet_case_2025} made distinction between time-shared and spatial-shared computing, arguing that \textit{``spatial-sharing is the norm... But time-sharing is the exception,''} and suggesting that \textit{``only time-sharing can truly mutualize all resources.''}  In their advocacy for a return to time-sharing, temporal flexibility or a  willingness to wait, is suggested as a primary mechanism for efficiency. However, they note along with Nystrom et al. \cite{nystrom_challenging_2021}that we might end up \textit{accepting longer wait times based on priorities, market-based rules,} raising the question of who can afford to move fast and who is forced to wait.
Other work problematizes attention toward personal will rather than system-level uses of time. McDonald et al., \cite{mcdonald_political_2017} for instance, engage questions of the political through their work on activism-based digital messaging tools, where issues of time, power, and technology intersect in further collective and political action. In these contexts, time becomes central to coordination, signaling when and how action can take place. In a similar vein, Blevis et al. propose a more structural and seemingly temporal intervention -- the cyclical cessation of digital infrastructure through observance a day of rest \cite{blevis_further_2017}. By framing downtime as a policy issue rather than an individual choice, they shift the burden of temporal management from user willpower to a collectively organized, infrastructural rhythm, thereby politicizing how and when digital systems should cease to operate. Snodgrass et al. \cite{snodgrass_designing_2024} invoke the permaculture principle of sustained observation before action, standing in opposition to ``agile'' development and rapid prototyping. This temporal framework requires understanding a site's specific rhythms (wind, sun, soil) as a prerequisite to technical intervention, effectively slowing down the design process to match ecological cycles. Leal et al. \cite{leal_hcis_2021} situate temporal dynamics within the Capitalocene, stating that \textit{``while historical colonialism has ended, its influence is ongoing as Coloniality\ldots a sister term to the Capitalocene as they refer to the same time.''} Their framework disrupts the linear progress narrative, suggesting that developments in computation are not moving society forward but are re-enacting historical extraction in real time.

\subsection{Biological and Ecological Time}

Attempts to synchronize computing with the non-linear, cyclical, and intermittent rhythms of the natural world we're noted in some contributions. Writing on strategies of degrowth, Sutherland \cite{sutherland_strategies_2022}  speaks of certain computational activities as what might be considered second-tier tasks and how they might be relegated to processes dependent on variability of wind energy.  Soden et al. \cite{soden_photovoltaic_2022} argue for \textit{``thinking of non-abundance of energy... not as a response to a crisis... but as a steady-state for future design.''} The assumption here is that a steady-state is one dictated by the sun and other biological and ecological rhythms.  Brain et al.\cite{brain_solar_2022} describe\textit{``follow-the-sun computing,''} which uses sunlight as a guiding principle for determining where certain computational tasks occur. By \textit{``automating decisions according to environmental dynamics,''} their experimental web platform envisions a kind of natural intelligence. A similar line of thinking is shown by Abbing \cite{abbing_this_2021} in their work on designing a low tech website powered by off-grid solar energy.  Colliaux et al. \cite{colliaux_computational_2022} proposed computation as a way to mediate and coordinate diverse ecological and biological temporalities. Their work suggests that the complexity of sustainable polyculture, characterized by overlapping and asynchronous cycles (e.g., hundreds of agricultural plant batches with distinct growth rates), exceeds unassisted human cognitive capacity, positioning computation as necessary for managing this temporal diversity.

\subsection{Afterlife and Waste Temporality}

Work in this category confronts time is computing as material permanence and the persistence of materials long after their utility has ended.  Michel and Frenkiel \cite{michel_ict_2025} deploy the concept of \textit{Zombie Technology}, stating: \textit{``These technologies are doomed to survive in a degraded form for a very long time.''} This creates a negative commons that is inherited, framing sustainability not as innovation but as the management of a toxic past. To address this, they propose the inverse legacy problem, noting that \textbf{}\textit{``whereas regular obsolescence is a forward-looking movement\ldots inverse legacy is a backward-facing movement.''}
Comber and Eriksson \cite{comber_computing_2023} extend the timescale of computing harms from the foreseeable present to generational and even geological temporal frames , fearing that\textit{``we look back and rename this phase of earth’s existence from Anthropocene to computocene.''} Their work understands \textit{``long-term''} through ecocide: {``damage which is irreversible or which cannot be redressed through natural recovery within a reasonable period of time.''} They suggest that computing-related waste is an enduring geological marker that outlives human lifetimes.

The contribution types identified in our review show that even when time is not explicitly engaged with, its assumptions and conceptualizations surface through metaphors and ways contributions are framed. In the next section, we discuss these findings alongside the author reflection and cross-disciplinary scholarship that has long treated time as an analytic concern.

\section{Discussion}
Stepping back, what do we learn by placing a decade of LIMITS scholarship alongside a personal experiment? We began with one author’s attempt to live within another temporal framework, an experience that was far from seamless. Rather than integrating smoothly, it produced moments of disorientation and discombobulation. These experiences are closely tied to the motivations that brought us to engage with the LIMITS community and shaped how we approached both the review and the interpretation of our findings.

We do not claim to have exhausted the many ways time and temporality can be understood. Temporalities are vast and varied, and the LIMITS community reflects this diversity through its interdisciplinary composition. While this diversity is a strength, it also raises the challenge of developing a shared language around time. At the same time, this may not be necessary. In other fields, time is not treated as a singular or uniform concept, but rather as a unit of analysis through which broader issues are examined \cite{Edensor2020, bear_time_2016, may_timespace_2003,marquardt_making_2021, gokmenoglu_temporality_2022,fent_political_2020}.

Our review suggests that many of the concerns central to the LIMITS community, such as sustainability, climate change, and long-term futures, are fundamentally temporal in nature. However, without explicitly attending to time as a unit of analysis, there is a risk of reproducing practices that take temporal assumptions for granted. Moments of discombobulation, as shared in the author reflection, point to these assumptions and signal opportunities for critically engaging with them. There is also a risk that, without explicitly attending to time, the viability of the work we do becomes tied to particular temporal assumptions about the future. In such cases, time is not neutral but actively shapes the conditions under which our work becomes relevant. These temporal framings often assume trajectories of worsening, where conditions are expected to deteriorate, positioning LIMITS work as necessary within those very futures. In this sense, temporal assumptions do not simply describe the future. We propose that they help produce it by shaping the conditions of possibility through which problems are defined, action is organized, and certain futures become thinkable while others are foreclosed. Rather than questioning these conditions, our work risks operating within them, responding to decline rather than interrogating the temporal understandings that make such decline appear inevitable. Attending to time as a unit of analysis therefore becomes crucial for making visible these conditions and for opening up alternative temporal frameworks through which different futures might be imagined.

Across the literature, the concept of a limit in computing emerges as inherently temporal. LIMITS scholarship engages with the chrononormativity of computing \cite{rifkin_beyond_2017}, including assumptions that faster is better, availability must be constant, and the future is a linear extension of the present. Yet time can also be mobilized as a site of resistance, where alternative temporal frameworks create openings for re-imagining how we design, build, and engage with computing systems. The personal experiment with the Chinese Diaspora Lunar Calendar illustrates this point where the experience of disorientation was an articulation of the work that temporal assumptions do. It made visible the tension between dominant and alternative temporal frameworks. Similarly, approaches such as backcasting \cite{pirson_notitle_2023} and speculative design \cite{bhardwaj_limits_2024} attempt to bridge present actions and long-term futures, but often do so without explicitly engaging the temporal assumptions they rely on. As Fent \cite{fent_political_2020}  and Paprocki \cite{Paprocki10112022} argue, practices of anticipation also carry the power to foreclose certain futures. Time, then, becomes central to to both constraint and possibility. Hendry \cite{hendry_transition_2021} describes LIMITS as a transition system shaping researchers, educators, and activists. Given this role, the community must be attentive to how it engages with time in both its research and practices. A step forward is to learn from traditions that have long resisted dominant temporal frameworks. For instance, Morosin \cite{morosin_comunalidad_2020} shows how Indigenous cosmovisions and values rooted in relationships between past, present, and future mobilize temporal politics in the Isthmus of Tehuantepec to challenge and slow the commodification of land and nature. Awâsis \cite{awasis_anishinaabe_2020} similarly describes Anishinaabe temporalities as relational and not reducible to the 24-hour clock. Taken together, these insights suggest that making time explicit as a unit of analysis is not peripheral to the work of LIMITS, but central to its commitments. Doing so allows the community to more fully engage with the temporal dimensions of sustainability, limits, and futures, and to reflect on the assumptions that shape what is considered possible.

\section{Conclusion}

Our work began by exploring non-chrononormative approaches to time. We then extended this inquiry to a review of a decade of LIMITS scholarship. We set out to explore how time and temporal frameworks are conceived within a community defined by limits and identified five types of temporal engagement. We found that even when time was not explicitly discussed or central to analysis, it appeared through underlying temporal assumptions, metaphors, and framings across LIMITS contributions. Revisiting the author reflections alongside cross-disciplinary scholarship that considers time as unit of analysis, we recognize that attending to and working through temporality can itself be a disorienting, even ``discombobulating,'' process, yet a necessary one given the temporal concerns at the center of LIMITS work, both in the causes of these concerns and in their potential solutions. We invite the LIMITS community to intentionally sit with these ``disorientations'' and ``discombobulations'' of temporalities and to more critically recognize and engage the temporal assumptions shaping its work.

\bibliographystyle{ACM-Reference-Format}
\bibliography{references, manual-add}

@inproceedings{pargman2023exploring,
  title={Exploring inner transition: Expanding computing for sustainability},
  author={Pargman, Daniel and Eriksson, Elina},
  booktitle={Proc. Workshop Comput. Limits},
  year={2023}
}

@article{Edensor2020,
title = {Time, temporality and environmental change},
journal = {Geoforum},
volume = {108},
pages = {255-258},
year = {2020},
issn = {0016-7185},
doi = {https://doi.org/10.1016/j.geoforum.2019.11.003},
url = {https://www.sciencedirect.com/science/article/pii/S0016718519303124},
author = {Tim Edensor and Lesley Head and Uma Kothari}
}

@inproceedings{wieczorek2026whose,
  title={Whose Time Counts? Temporal Arrangements in Sociotechnical Infrastructures},
  author={Wieczorek, Catherine and Tran, Anh-Ton and Lin, Cindy Kaiying and Forlano, Laura and DiSalvo, Carl and Bardzell, Shaowen},
  booktitle={Proceedings of the 2026 CHI Conference on Human Factors in Computing Systems},
  pages={1--18},
  year={2026}
}

@article{Paprocki10112022,
author = {Kasia Paprocki},
title = {Anticipatory ruination},
journal = {The Journal of Peasant Studies},
volume = {49},
number = {7},
pages = {1399--1408},
year = {2022},
publisher = {Routledge},
doi = {10.1080/03066150.2022.2113068},
}

@incollection{hartmann_normative_2019,
	address = {Cham},
	title = {The {Normative} {Framework} of ({Mobile}) {Time}: {Chrononormativity}, {Power}-{Chronography}, and {Mobilities}},
	isbn = {978-3-030-24949-6 978-3-030-24950-2},
	shorttitle = {The {Normative} {Framework} of ({Mobile}) {Time}},
	url = {http://link.springer.com/10.1007/978-3-030-24950-2_3},
	doi = {10.1007/978-3-030-24950-2_3},
	language = {en},
	urldate = {2026-05-28},
	booktitle = {Mediated {Time}},
	publisher = {Springer International Publishing},
	author = {Hartmann, Maren},
	editor = {Hartmann, Maren and Prommer, Elizabeth and Deckner, Karin and Görland, Stephan O.},
	year = {2019},
	pages = {45--65},
}

@article{samuels_six_2017,
	title = {Six {Ways} of {Looking} at {Crip} {Time}},
	volume = {37},
	issn = {2159-8371},
	url = {https://dsq-sds.org/article/id/1565/},
	doi = {10.18061/dsq.v37i3.5824},
	language = {None},
	number = {3},
	urldate = {2026-05-28},
	journal = {Disability Studies Quarterly},
	publisher = {The Ohio State University Libraries},
	author = {Samuels, Ellen},
	month = aug,
	year = {2017},
}

@article{bear_time_2016,
	title = {Time as {Technique}},
	volume = {45},
	issn = {0084-6570, 1545-4290},
	url = {https://www.annualreviews.org/doi/10.1146/annurev-anthro-102313-030159},
	doi = {10.1146/annurev-anthro-102313-030159},
	language = {en},
	number = {1},
	urldate = {2026-04-04},
	journal = {Annual Review of Anthropology},
	author = {Bear, Laura},
	month = oct,
	year = {2016},
	pages = {487--502},
}

@book{may_timespace_2003,
	address = {London},
	title = {Timespace: {Geographies} of {Temporality}},
	isbn = {978-0-203-36067-5},
	shorttitle = {Timespace},
	doi = {10.4324/9780203360675},
	publisher = {Routledge},
	editor = {May, Jon and Thrift, Nigel},
	month = aug,
	year = {2003},
}

@article{marquardt_making_2021,
	title = {Making time, making politics: {Problematizing} temporality in energy and climate studies},
	volume = {76},
	issn = {2214-6296},
	shorttitle = {Making time, making politics},
	url = {https://www.sciencedirect.com/science/article/pii/S2214629621001663},
	doi = {10.1016/j.erss.2021.102073},
	urldate = {2026-04-03},
	journal = {Energy Research \& Social Science},
	author = {Marquardt, Jens and Delina, Laurence L.},
	month = jun,
	year = {2021},
	pages = {102073},
}

@article{gokmenoglu_temporality_2022,
	title = {Temporality in the social sciences: {New} directions for a political sociology of time},
	volume = {73},
	issn = {0007-1315},
	shorttitle = {Temporality in the social sciences},
	url = {https://pmc.ncbi.nlm.nih.gov/articles/PMC9314931/},
	doi = {10.1111/1468-4446.12938},
	number = {3},
	urldate = {2026-04-03},
	journal = {The British Journal of Sociology},
	author = {Gokmenoglu, Birgan},
	month = jun,
	year = {2022},
	pages = {643--653},
}

@inproceedings{pirson_notitle_2023,
	address = {Virtual},
	url = {https://limits.pubpub.org/pub/8ld7lmdf},
	doi = {10.21428/bf6fb269.6af396ff},
	language = {en},
	urldate = {2025-11-03},
	booktitle = {Ninth {Computing} within {Limits} 2023},
	publisher = {LIMITS},
	author = {Pirson, Thibault and Golard, Louis and Bol, David},
	month = jun,
	year = {2023},
}

@inproceedings{brain_solar_2022,
	title = {Solar {Protocol}: {Exploring} {Energy}-{Centered} {Design}},
	language = {en},
	author = {Brain, Tega and Nathanson, Alex and Piantella, Benedetta},
	year = {2022},
}

@article{morosin_comunalidad_2020,
	title = {Comunalidad, {Guendaliza}'a and anti-mine mobilizations in the {Isthmus} of {Tehuantepec}},
	volume = {27},
	issn = {1073-0451},
	url = {http://journals.librarypublishing.arizona.edu/jpe/article/id/2256/},
	doi = {10.2458/v27i1.23237},
	language = {None},
	number = {1},
	urldate = {2025-12-12},
	journal = {Journal of Political Ecology},
	publisher = {University of Arizona Libraries},
	author = {Morosin, Alessandro},
	month = jan,
	year = {2020},
}

@article{awasis_anishinaabe_2020,
	title = {"{Anishinaabe} time": temporalities and impact assessment in pipeline reviews},
	volume = {27},
	issn = {1073-0451},
	shorttitle = {"{Anishinaabe} time"},
	url = {http://journals.librarypublishing.arizona.edu/jpe/article/id/2259/},
	doi = {10.2458/v27i1.23236},
	language = {None},
	number = {1},
	urldate = {2025-12-12},
	journal = {Journal of Political Ecology},
	publisher = {University of Arizona Libraries},
	author = {Awasis, Sakihitowin},
	month = jan,
	year = {2020},
}

@incollection{zhou_following_2023,
	address = {Cham},
	title = {Following the {Rhythm} of {Nature}: {Wisdom} in the 24 {Solar} {Terms} and the 12 {Constellations}},
	isbn = {978-3-031-17156-7 978-3-031-17157-4},
	shorttitle = {Following the {Rhythm} of {Nature}},
	url = {https://link.springer.com/10.1007/978-3-031-17157-4_2},
	doi = {10.1007/978-3-031-17157-4_2},
	language = {en},
	urldate = {2025-12-11},
	booktitle = {Science {Education} and {International} {Cross}-{Cultural} {Reciprocal} {Learning}},
	publisher = {Springer International Publishing},
	author = {Chen, Xue and Liao, Yanan and Li, Yuanrong},
	editor = {Zhou, George and Li, Yuanrong and Luo, Jian},
	year = {2023},
	pages = {11--26},
}

@inproceedings{sutherland_strategies_2022,
	address = {Virtual},
	title = {Strategies for {Degrowth} {Computing}},
	url = {https://limits.pubpub.org/pub/strat},
	doi = {10.21428/bf6fb269.04676652},
	language = {en},
	urldate = {2025-10-16},
	booktitle = {Eighth {Workshop} on {Computing} within {Limits} 2022},
	publisher = {LIMITS},
	author = {Sutherland, Brian},
	month = jun,
	year = {2022},
}

@inproceedings{abbing_this_2021,
	title = {‘{This} is a solar-powered website, which means it sometimes goes offline’: a design inquiry into degrowth and {ICT}},
	shorttitle = {‘{This} is a solar-powered website, which means it sometimes goes offline’},
	url = {https://limits.pubpub.org/pub/lecuxefc},
	doi = {10.21428/bf6fb269.e78d19f6},
	language = {en},
	urldate = {2025-11-03},
	booktitle = {{LIMITS} {Workshop} on {Computing} within {Limits}},
	author = {Abbing, Roel Roscam},
	month = jun,
	year = {2021},
}

@inproceedings{snodgrass_designing_2024,
	title = {Designing grid-liberated servers for regenerative energy communities},
	language = {en},
	booktitle = {{LIMITS} '},
	author = {Snodgrass, Eric and Pritchard, Helen and Moss, Miranda and Gustafsson, Daniel and Zapico, Jorge Luis},
	year = {2024},
}

@inproceedings{bhardwaj_pathways_2023,
	address = {Virtual},
	title = {Pathways to urban sustainability: {Design} perspectives on a data curation and visualization platform},
	shorttitle = {Pathways to urban sustainability},
	url = {https://limits.pubpub.org/pub/iw8v1y2m},
	doi = {10.21428/bf6fb269.541455de},
	language = {en},
	urldate = {2025-11-03},
	booktitle = {Ninth {Computing} within {Limits} 2023},
	publisher = {LIMITS},
	author = {Bhardwaj, Esta and Qiao, Han and Becker, Christoph},
	month = jun,
	year = {2023},
}

@inproceedings{colliaux_computational_2022,
	title = {Computational {Agroecology}: {Should} we bet the microfarm on it?},
	shorttitle = {Computational {Agroecology}},
	url = {https://limits.pubpub.org/pub/comput},
	doi = {10.21428/bf6fb269.5b8520b2},
	language = {en},
	urldate = {2025-11-03},
	booktitle = {Eighth {Workshop} on {Computing} within {Limits} 2022},
	publisher = {LIMITS},
	author = {Colliaux, David and Minchin, Jonathan and Goelzer, Sebastien and Hanappe, Peter},
	month = jun,
	year = {2022},
}

@inproceedings{pargman_limits_nodate,
	title = {On the {Limits} of {Limits}},
	language = {en},
	author = {Pargman, Daniel},
}

@inproceedings{michel_ict_2025,
	title = {{ICT} {Within} {Limits} {Is} {Bound} {To} {Be} {Old}-{Fashioned} {By} {Design}},
	language = {en},
	author = {Michel, Olivier and Frenkiel, Émilie},
	year = {2025},
}

@inproceedings{jacquet_case_2025,
	title = {The {Case} for {Time}-{Shared} {Computing} {Resources}},
	language = {en},
	author = {Jacquet, Pierre and Luxey-Bitri, Adrien},
	year = {2025},
}

@inproceedings{bhardwaj_limits_2024,
	title = {Limits at a {Distance}: {Design} {Directions} to {Address} {Psychological} {Distance} in {Policy} {Decisions} {Affecting} {Planetary} {Boundaries}},
	language = {en},
	author = {Bhardwaj, Eshta and Qiao, Han and Becker, Christoph},
	year = {2024},
}

@inproceedings{soden_photovoltaic_2022,
	title = {Photovoltaic {Imagination}: {Solar} {Strategies} for {Community} {Integrated} {Research} and {Graduate} {Training}},
	language = {en},
	author = {Soden, Robert and Ratto, Matt and Verhoeven, G Arno and Simon, Bart},
	year = {2022},
}

@inproceedings{penzenstadler_pursuit_2021,
	title = {The {Pursuit} of {Essence}: {Realizing} {Expansion} and {Oneness} by {Limitation}},
	shorttitle = {The {Pursuit} of {Essence}},
	url = {https://limits.pubpub.org/pub/f5x0ct9v},
	doi = {10.21428/bf6fb269.390c5acb},
	language = {en},
	urldate = {2025-11-03},
	booktitle = {{LIMITS} {Workshop} on {Computing} within {Limits}},
	author = {Penzenstadler, Birgit},
	month = jun,
	year = {2021},
}

@inproceedings{nystrom_challenging_2021,
	title = {Challenging the image of the altruistic and flexible household in the smart grid using design fiction},
	url = {https://limits.pubpub.org/pub/5mndg0b5},
	doi = {10.21428/bf6fb269.824814be},
	language = {en},
	urldate = {2025-11-03},
	booktitle = {{LIMITS} {Workshop} on {Computing} within {Limits}},
	author = {Nyström, Sofie and Rivera, Miriam Börjesson and Hedin, Björn and Katzeff, Cecilia and Menon, Arjun Rajendran},
	month = jun,
	year = {2021},
}

@inproceedings{mayhew_materiality_2021,
	title = {Materiality {Matters} in {Computing} {Education}: {A} {Duoethnography} of {Two} {Digital} {Logic} {Educators}},
	shorttitle = {Materiality {Matters} in {Computing} {Education}},
	url = {https://limits.pubpub.org/pub/pkxwkg1p},
	doi = {10.21428/bf6fb269.19c98f37},
	language = {en},
	urldate = {2025-11-03},
	booktitle = {{LIMITS} {Workshop} on {Computing} within {Limits}},
	author = {Mayhew, Eric J. and Patitsas, Elizabeth},
	month = jun,
	year = {2021},
}

@inproceedings{leal_hcis_2021,
	title = {{HCI}’s {Role} in the {Capitalocene}: {Lessons} {Learned} from an {HCI} {Master} {Course} {Across} the {Globe}},
	shorttitle = {{HCI}’s {Role} in the {Capitalocene}},
	url = {https://limits.pubpub.org/pub/f8ee6iac},
	doi = {10.21428/bf6fb269.67a8d057},
	language = {en},
	urldate = {2025-11-03},
	booktitle = {{LIMITS} {Workshop} on {Computing} within {Limits}},
	author = {Leal, Débora De Castro and Krüger, Max and Ahmadi, Michael and Appiah, Jason and Gómez, Ricardo A. Baquero and Courtney, Daniel and Daee, Ata and Ciciolli, María Belén Giménez and Hieber, Lena and Hossain, Md Shakhawat and Lee, Jeongmin and Plogmann, Ramona and Pinto, Liliana Savage and Sinnathurai, Sasmitha and Yepez, Darinka and Wulf, Volker},
	month = jun,
	year = {2021},
}

@inproceedings{hendry_transition_2021,
	title = {Transition {Discourse}, {Food}, and {Computing} within {Limits}},
	url = {https://limits.pubpub.org/pub/1bkgrc4d},
	doi = {10.21428/bf6fb269.244d268b},
	language = {en},
	urldate = {2025-11-03},
	booktitle = {{LIMITS} {Workshop} on {Computing} within {Limits}},
	author = {Hendry, David G.},
	month = jun,
	year = {2021},
}

@inproceedings{hilty_computing_nodate,
	title = {Computing {Efficiency}, {Sufficiency}, and {Self}-sufficiency: {A} {Model} for {Sustainability}?},
	language = {en},
	author = {Hilty, Lorenz M},
}

@inproceedings{penzenstadler_leverage_2020,
	address = {New York, NY, USA},
	series = {{ICT4S2020}},
	title = {Leverage {Points} for {Focus} {Flow} and {Communitas}},
	isbn = {978-1-4503-7595-5},
	url = {https://dl.acm.org/doi/10.1145/3401335.3401826},
	doi = {10.1145/3401335.3401826},
	urldate = {2025-11-03},
	booktitle = {Proceedings of the 7th {International} {Conference} on {ICT} for {Sustainability}},
	publisher = {Association for Computing Machinery},
	author = {Penzenstadler, Birgit},
	month = jul,
	year = {2020},
	pages = {267--274},
}

@inproceedings{comber_computing_2023,
	address = {Virtual},
	title = {Computing as {Ecocide}},
	url = {https://limits.pubpub.org/pub/a8h46wqy},
	doi = {10.21428/bf6fb269.9fcdd0c0},
	language = {en},
	urldate = {2025-11-03},
	booktitle = {Ninth {Computing} within {Limits} 2023},
	publisher = {LIMITS},
	author = {Comber, Rob and Eriksson, Elina},
	month = jun,
	year = {2023},
}

@inproceedings{pargman_experimenting_2019,
	address = {Lappeenranta Finland},
	title = {Experimenting with {Novel} {Forms} of {Computing}: {The} case of the {Swedish} {Citizen} {Observatory} for {Water} {Quality} {Conservation}},
	isbn = {978-1-4503-7281-7},
	shorttitle = {Experimenting with {Novel} {Forms} of {Computing}},
	url = {https://dl.acm.org/doi/10.1145/3338103.3338111},
	doi = {10.1145/3338103.3338111},
	language = {en},
	urldate = {2025-11-03},
	booktitle = {Proceedings of the {Fifth} {Workshop} on {Computing} within {Limits}},
	publisher = {ACM},
	author = {Pargman, Teresa Cerratto and Joshi, Somya and Wehn, Uta},
	month = jun,
	year = {2019},
	pages = {1--10},
}

@inproceedings{mann_regenerative_2018,
	address = {Toronto Ontario Canada},
	title = {Regenerative computing: de-limiting hope},
	isbn = {978-1-4503-6575-8},
	shorttitle = {Regenerative computing},
	url = {https://dl.acm.org/doi/10.1145/3232617.3232618},
	doi = {10.1145/3232617.3232618},
	language = {en},
	urldate = {2025-11-03},
	booktitle = {Proceedings of the 2018 {Workshop} on {Computing} within {Limits}},
	publisher = {ACM},
	author = {Mann, Samuel and Bates, Oliver and Forsyth, Glenys and Osborne, Phil},
	month = may,
	year = {2018},
	pages = {1--10},
}

@inproceedings{eriksson_meeting_2018,
	address = {Toronto Ontario Canada},
	title = {Meeting the future in the past - using counterfactual history to imagine computing futures},
	isbn = {978-1-4503-6575-8},
	url = {https://dl.acm.org/doi/10.1145/3232617.3232621},
	doi = {10.1145/3232617.3232621},
	language = {en},
	urldate = {2025-11-03},
	booktitle = {Proceedings of the 2018 {Workshop} on {Computing} within {Limits}},
	publisher = {ACM},
	author = {Eriksson, Elina and Pargman, Daniel},
	month = may,
	year = {2018},
	pages = {1--8},
}

@inproceedings{mcdonald_political_2017,
	address = {Santa Barbara California USA},
	title = {Political {Realities} of {Digital} {Communication}: {The} {Limits} of {Value} from {Digital} {Messages} to {Members} of the {US} {Congress}},
	isbn = {978-1-4503-4950-5},
	shorttitle = {Political {Realities} of {Digital} {Communication}},
	url = {https://dl.acm.org/doi/10.1145/3080556.3080565},
	doi = {10.1145/3080556.3080565},
	language = {en},
	urldate = {2025-11-03},
	booktitle = {Proceedings of the 2017 {Workshop} on {Computing} {Within} {Limits}},
	publisher = {ACM},
	author = {McDonald, Samantha and Nardi, Bonnie and Tomlinson, Bill},
	month = jun,
	year = {2017},
	pages = {129--138},
}

@inproceedings{blevis_further_2017,
	address = {Santa Barbara California USA},
	title = {Further {Connecting} {Sustainable} {Interaction} {Design} with {Sustainable} {Digital} {Infrastructure} {Design}},
	isbn = {978-1-4503-4950-5},
	url = {https://dl.acm.org/doi/10.1145/3080556.3080568},
	doi = {10.1145/3080556.3080568},
	language = {en},
	urldate = {2025-11-03},
	booktitle = {Proceedings of the 2017 {Workshop} on {Computing} {Within} {Limits}},
	publisher = {ACM},
	author = {Blevis, Eli and Preist, Chris and Schien, Daniel and Ho, Priscilla},
	month = jun,
	year = {2017},
	pages = {71--83},
}

@article{caseldine_conceptions_2012,
	title = {Conceptions of time in (paleo)climate science and some implications},
	volume = {3},
	copyright = {http://onlinelibrary.wiley.com/termsAndConditions\#vor},
	issn = {1757-7780, 1757-7799},
	url = {https://wires.onlinelibrary.wiley.com/doi/10.1002/wcc.178},
	doi = {10.1002/wcc.178},
	language = {en},
	number = {4},
	urldate = {2025-10-26},
	journal = {WIREs Climate Change},
	author = {Caseldine, Chris},
	month = jul,
	year = {2012},
	pages = {329--338},
}

@article{iparraguirre_time_2015,
	title = {Time, temporality and cultural rhythmics: {An} anthropological case study},
	shorttitle = {Time, temporality and cultural rhythmics},
	url = {https://journals.sagepub.com/doi/full/10.1177/0961463X15579802?casa_token=7e56pKGzR1wAAAAA%3AcF3zz86pm1w7B56YA30McywdYOR4IN9OW4NmPV-213GX3ieJWPr7xRyliFcvqnAOiG-ohJEGOTII},
	language = {en},
	urldate = {2025-10-25},
	journal = {Time \& Society},
	author = {Iparraguirre, Gonzalo},
	month = apr,
	year = {2015},
}

@article{irvine_deep_2014,
	title = {Deep time: an anthropological problem:},
	volume = {22},
	copyright = {http://doi.wiley.com/10.1002/tdm\_license\_1.1},
	issn = {09640282},
	shorttitle = {Deep time},
	url = {http://berghahnjournals.com/view/journals/saas/22/2/soca12067.xml},
	doi = {10.1111/1469-8676.12067},
	language = {en},
	number = {2},
	urldate = {2025-10-26},
	journal = {Social Anthropology},
	author = {Irvine, Richard},
	month = may,
	year = {2014},
	pages = {157--172},
}

@article{wiberg_time_2021,
	title = {Time and {Temporality} in {HCI} {Research}},
	volume = {33},
	issn = {1873-7951},
	url = {https://ieeexplore.ieee.org/abstract/document/9646577},
	doi = {10.1093/iwc/iwab025},
	number = {3},
	urldate = {2025-10-26},
	journal = {Interacting with Computers},
	author = {Wiberg, Mikael and Stolterman, Erik},
	month = sep,
	year = {2021},
	pages = {250--270},
}

@article{kafer_after_2021,
	title = {After {Crip}, {Crip} {Afters}},
	volume = {120},
	issn = {0038-2876, 1527-8026},
	url = {https://read.dukeupress.edu/south-atlantic-quarterly/article/120/2/415/173312/After-Crip-Crip-Afters},
	doi = {10.1215/00382876-8916158},
	language = {en},
	number = {2},
	urldate = {2024-12-02},
	journal = {South Atlantic Quarterly},
	author = {Kafer, Alison},
	month = apr,
	year = {2021},
	pages = {415--434},
}

@book{rifkin_beyond_2017,
	address = {Durham London},
	title = {Beyond settler time: temporal sovereignty and indigenous self-determination},
	isbn = {978-0-8223-6297-5 978-0-8223-6285-2},
	shorttitle = {Beyond settler time},
	language = {eng},
	publisher = {Duke University Press},
	author = {Rifkin, Mark},
	year = {2017},
}

@article{fent_political_2020,
	title = {Political ecologies of time and temporality in resource extraction},
	volume = {27},
	copyright = {Copyright (c) 2020 Ashley Fent, Erik Kojola},
	issn = {1073-0451},
	url = {https://journals.uair.arizona.edu/index.php/JPE/article/view/23252},
	language = {en},
	number = {1},
	urldate = {2025-10-25},
	journal = {Journal of Political Ecology},
	author = {Fent, Ashley and Kojola, Erik},
	month = nov,
	year = {2020},
	pages = {819--829},
}

@article{rapp_introduction_2022,
	title = {Introduction to the special issue on time and {HCI}},
	volume = {37},
	issn = {0737-0024},
	url = {https://doi.org/10.1080/07370024.2021.1955681},
	doi = {10.1080/07370024.2021.1955681},
	number = {1},
	urldate = {2025-10-23},
	journal = {Human–Computer Interaction},
	publisher = {Taylor \& Francis},
	author = {Rapp, Amon and Odom, William and Pschetz, Larissa and Petrelli, Daniela},
	month = jan,
	year = {2022},
	note = {\_eprint: https://doi.org/10.1080/07370024.2021.1955681},
	pages = {1--14},
}

@article{senda-cook_engaging_2023,
	title = {Engaging complex temporalities in environmental rhetoric},
	volume = {8},
	issn = {2297-900X},
	url = {https://www.frontiersin.org/journals/communication/articles/10.3389/fcomm.2023.1176887/full},
	doi = {10.3389/fcomm.2023.1176887},
	language = {English},
	urldate = {2025-10-23},
	journal = {Frontiers in Communication},
	publisher = {Frontiers},
	author = {Senda-Cook, Samantha and Endres, Danielle and Sowards, Stacey K. and McGreavy, Bridie},
	month = may,
	year = {2023},
}

@article{zohar_when_2024,
	title = {When we talk about time, we mean many different things: employing visual mapping to think through more-than-human temporalities in participatory design},
	volume = {9},
	issn = {2297-900X},
	shorttitle = {When we talk about time, we mean many different things},
	url = {https://www.frontiersin.org/journals/communication/articles/10.3389/fcomm.2024.1440257/full},
	doi = {10.3389/fcomm.2024.1440257},
	language = {English},
	urldate = {2025-10-23},
	journal = {Frontiers in Communication},
	publisher = {Frontiers},
	author = {Zohar, Hadas and Simeone, Luca and de Götzen, Amalia and Morelli, Nicola},
	month = aug,
	year = {2024},
}

@article{rahm-skageby_hci_2022,
	title = {{HCI} and deep time: toward deep time design thinking},
	volume = {37},
	issn = {0737-0024},
	shorttitle = {{HCI} and deep time},
	url = {https://doi.org/10.1080/07370024.2021.1902328},
	doi = {10.1080/07370024.2021.1902328},
	number = {1},
	urldate = {2025-10-23},
	journal = {Human–Computer Interaction},
	publisher = {Taylor \& Francis},
	author = {Rahm-Skågeby, Jörgen and Rahm, Lina},
	month = jan,
	year = {2022},
	note = {\_eprint: https://doi.org/10.1080/07370024.2021.1902328},
	pages = {15--28},
}

@article{griffiths_time_2023,
	title = {Time and the {Anthropocene}: {Making} more-than-human temporalities legible through environmental observations and creative methods},
	volume = {32},
	issn = {0961-463X},
	url = {https://doi.org/10.1177/0961463X231202928},
	doi = {10.1177/0961463X231202928},
	number = {4},
	urldate = {2025-10-23},
	journal = {Time \& Society},
	publisher = {SAGE Publications Ltd},
	author = {Griffiths, Rupert},
	month = nov,
	year = {2023},
	pages = {461--487},
}

@inproceedings{laurell_thorslund_exploring_2025,
	address = {Yokohama Japan},
	title = {Exploring {Assumptions} about {Sustainability}: {Towards} a {Constructive} {Framework} for {Action} in {Sustainable} {HCI}},
	isbn = {979-8-4007-1394-1},
	shorttitle = {Exploring {Assumptions} about {Sustainability}},
	url = {https://dl.acm.org/doi/10.1145/3706598.3714001},
	doi = {10.1145/3706598.3714001},
	language = {en},
	urldate = {2025-10-13},
	booktitle = {Proceedings of the 2025 {CHI} {Conference} on {Human} {Factors} in {Computing} {Systems}},
	publisher = {ACM},
	author = {Laurell Thorslund, Minna and Leifler, Ola},
	month = apr,
	year = {2025},
	pages = {1--13},
}

@book{mbiti_african_1970,
	title = {African religions and philosophy},
	isbn = {978-0-385-03713-6},
	url = {http://archive.org/details/africanreligions00john},
	language = {eng},
	urldate = {2025-10-16},
	publisher = {Garden City, N.Y. : Anchor Books},
	author = {Mbiti, John S.},
	collaborator = {{Internet Archive}},
	year = {1970},
}

\appendix

\end{document}